\documentclass[aps,prd,nofootinbib,twocolumn,superscriptaddress,preprintnumbers,balancelastpage,longbibliography]{revtex4-2}

\usepackage{placeins}
\usepackage{amsmath,amssymb,mathtools,bm}
\usepackage{graphicx, color, hepunits}
\usepackage[dvipsnames]{xcolor}
\usepackage{cancel}
\usepackage[normalem]{ulem}
\usepackage{float}
\usepackage{filecontents}
\usepackage{multirow}
\usepackage{tikz}

\usetikzlibrary{decorations.pathmorphing}
\usetikzlibrary{arrows}
\usepackage{circuitikz}
\usetikzlibrary{arrows.meta}
\usetikzlibrary{shapes.misc}
\usetikzlibrary{positioning}
\usetikzlibrary{decorations.markings}

\usepackage[compat=1.1.0]{tikz-feynman}

\usepackage{hyperref} %Automatically links \label and \ref commands; Always load last
\hypersetup{
    colorlinks=true,       % false: boxed links; true: colored links
    linkcolor=blue,        % color of internal links
    citecolor=blue,        % color of links to bibliography
    filecolor=magenta,     % color of file links
    urlcolor=blue          % color of external links
}
\usepackage[utf8]{inputenc}
\usepackage[english]{babel}

\usetikzlibrary{shapes.misc}
\tikzset{cross/.style={cross out, draw=black, minimum size=2*(#1-\pgflinewidth), inner sep=0pt, outer sep=0pt},
cross/.default={1pt}}

\newcommand{\ep}[1]{\epsilon_{#1}}

\begin{document}

\title{Strong CP and Flavor in Multi-Higgs Theories}

\author{Lawrence Hall}
\affiliation{Berkeley Center for Theoretical Physics, University of California, Berkeley, CA 94720, U.S.A.}
\affiliation{Theoretical Physics Group, Lawrence Berkeley National Laboratory, Berkeley, CA 94720, U.S.A.}

\author{Claudio Andrea Manzari}
\affiliation{Berkeley Center for Theoretical Physics, University of California, Berkeley, CA 94720, U.S.A.}
\affiliation{Theoretical Physics Group, Lawrence Berkeley National Laboratory, Berkeley, CA 94720, U.S.A.}

\author{Bea Noether}
\affiliation{Berkeley Center for Theoretical Physics, University of California, Berkeley, CA 94720, U.S.A.}
\affiliation{Theoretical Physics Group, Lawrence Berkeley National Laboratory, Berkeley, CA 94720, U.S.A.}

%\date{\today}

\begin{abstract}
We introduce a class of multi-Higgs doublet extensions of the Standard Model that solve the strong CP problem with profound consequences for the flavor sector. The Yukawa matrices are constrained to have many zero entries by a ``Higgs-Flavor" symmetry, $G_{\rm HF}$, that acts on Higgs and quark fields. The violation of both CP and $G_{\rm HF}$ occurs in the Higgs mass matrix so that, for certain choices of $G_{\rm HF}$ charges, the strong CP parameter $\bar{\theta}$ is zero at tree-level. Radiative corrections to $\bar{\theta}$ are computed in this class of theories. They vanish in  realistic two-Higgs doublet models with $G_{\rm HF} = \mathbb{Z}_3$. We also construct realistic  three-Higgs models with $G_{\rm HF} = \rm U(1)$, where the one-loop results for $\bar{\theta}$ are model-dependent. Requiring $\bar{\theta}< 10^{-10}$ has important implications for the flavor problem by constraining the Yukawa coupling and Higgs mass matrices.
Contributions to $\bar{\theta}$ from higher-dimension operators are computed at 1-loop and can also be sufficiently small, although the hierarchy problem of this class of theories is worse than in the Standard Model.
\end{abstract}
\maketitle

The violation of Parity (P) and its combination with Charge-conjugation (CP) in the strong interactions, described by the parameter $\bar{\theta}$, are constrained by the experimental limit on the neutron electric dipole moment (nEDM): $\bar{\theta} \leq 10^{-10}$~\cite{Pendlebury:2015lrz}. Hence $\bar{\theta}$ is the smallest dimensionless parameter of the Standard Model (SM). This small value is particularly puzzling because of the lack of theoretical understanding of it, and because the dimensionless parameter describing CP violation in the weak interactions is order unity. 
Within the SM, the parameter $\bar{\theta}$ has two sources. Firstly, the Lagrangian of Quantum Chromodynamics (QCD), restricted just by Lorentz and gauge invariance,  contains a CP-odd term, $\theta_{\rm QCD}\, g_s^2/(32\pi^2)\, G_a^{\mu\nu}\tilde{G}_{a,\mu\nu}$, where $G_a^{\mu\nu}$ is the gluon field strength tensor, $g_s$ the strong coupling constant, and $\tilde{G}_{a,\mu\nu} \equiv \frac{1}{2}\epsilon_{\mu\nu\alpha\beta}G_a^{\alpha\beta}$. Moreover, $\bar\theta$ receives a contribution through the chiral transformation needed to diagonalize the quarks Yukawa matrices $y^u$ and $y^d$,
\begin{align}
\bar{\theta}=\theta_\text{QCD} +\arg\det(y^uy^d)
\,.
\label{eq:thetaterm2}
\end{align}
The two contributions to $\bar{\theta}$ arise from very different physics and have no reason to cancel.

% In the PQ solution~\cite{Peccei:1977hh,Peccei:1977ur} to this so-called strong CP problem, the spontaneous breaking of an anomalous global $\rm U(1)$ symmetry leads to a very light pseudoscalar field~\cite{Weinberg:1977ma,Wilczek:1977pj} that dynamically relaxes $\bar{\theta}$ to zero at the cosmological era of the QCD phase transition. This scenario has received great attention because the resulting axion particle can be dark matter and offers a wide range of experimental probes. 

Solutions based on an underlying CP or P symmetry at high energies that is softly or spontaneously broken have profound consequences for the flavor sector of the SM. In these theories, the bare parameter $\theta_{QCD}$ in Eq.~\ref{eq:thetaterm2} is forced to be zero by either CP or P, and one must explain why $\arg\det(y^uy^d)$ is so small even though the order unity phase of the observed CP violation in weak interactions also arises from the quark mass matrices.

In the case of CP, Nelson and Barr~\cite{Nelson:1983zb,Barr:1984qx} introduced $N_Q$ extra generations of heavy vector-like quarks, $Q$, that mix with the SM quarks, $q$, through the interactions with new scalars, so that the Yukawa interactions are now described by square $3 + N_Q$ dimensional matrices. Careful model building arranges for the $Qq$ block of these matrices to vanish, while all CP-violating parameters appear in the $qQ$ block. The observed CP violation of the weak interactions arises from the mixing between $Q$ and $q$. As well as being enlarged, the Yukawa matrices involve more parameters than in the SM.  Different approaches with spontaneously CP breaking have been proposed that make use of the non-renormalization theorems of supersymmetry~\cite{Hiller:2002um,Feruglio:2023uof, Feruglio:2024ytl, Penedo:2024gtb}.

There is another scheme based on soft or spontaneous breaking of CP symmetry; it predates the Nelson-Barr framework, but has not been widely studied and appears to be largely forgotten. Instead of adding additional quark generations with CP violation in the mixing with SM quarks, it augments the SM by having more than one copy of the Higgs doublet, $\phi \rightarrow \phi_\alpha$. CP violation appears only in the Higgs mass matrix. The quark and Higgs fields transform under a flavor symmetry in such a way that the Yukawa matrices are highly restricted: each entry in the quark mass matrices is either zero or arises from a single Higgs field.  Hypercharge gauge symmetry forces all Yukawa interactions of the up (down) quarks to involve $\phi_\alpha$ ($\phi_\alpha^*$).  If the determinant of the up quark mass matrix is proportional to a single product of $\phi_a$ fields, $\Pi(\phi_\alpha)$, and if the same product appears in the determinant of the down quark masses, $\Pi(\phi_\alpha^*)$,  then $\arg\det{y_uy_d} = 0$ by a cancellation between up and down sectors. 

The hierarchy problem of the SM will not be made significantly worse if the effective field theory with multiple Higgs doublets is valid over a small energy interval. As in Nelson-Barr theories, one must ensure that the radiative contributions to $\bar{\theta}$ are sufficiently small.

Georgi constructed the first model of this type in 1978~\cite{Georgi:1978xz}. A 3-generation $SU(5)$ grand unified theory, with flavor group $\mathbb{Z}_{16}$, was constructed in 1979 by Mohapatra and Wyler~\cite{Mohapatra:1979kg}, and an extension of the SM at the weak scale was constructed by Glashow in 2001~\cite{Glashow:2001yz}. In light of modern data, only Glashow's model is phenomenologically viable, and we will show that it does does not convincingly solve the strong CP problem.

In this letter we provide a general framework for multi-Higgs extensions of the SM that solve the strong CP problem at tree-level by a combination of a flavor symmetry that strongly restricts the form of Yukawa interactions, together with soft or spontaneous breaking of this symmetry and of CP in the Higgs mass matrix.

\vspace{0.2cm}
{\bf Framework.---}
The Standard Model is minimally extended by increasing the number of Higgs doublets from 1 to $N$, all with the same quantum numbers. There are no additional fermions and no extension to the gauge group. All but one of the Higgs doublets acquires a mass of order $M$, and below $M$ the minimal SM is the effective field theory (EFT) describing the model. The lower bound on $M$ arises from limits on FCNC and from direct searches for scalar mesons at the LHC, and is typically expected to be about a TeV. We assume that the heavy Higgs bosons obey these mass bounds. There is no upper bound, although the hierarchy problem of the SM becomes more severe as $M$ is increased. The UV completion of the multi-Higgs theory occurs at a scale $\Lambda$. For the hierarchy problem of the multi-Higgs theory to be mild, $\Lambda$ should not be too far above $M$. 
\begin{figure}[tb]
\centering
\begin{tikzpicture}
\draw[solid,thick, -{Stealth[black]}] (0,0) -- (0,7) node[above] {$E$};
\draw[thick] (0.2,4) -- ++ (-0.4,0.) node[left, xshift=-0.3cm] {$M$};
\draw[dashed, thick] (0.2,4) -- (6.,4);
\draw[thick] (0.2,3) -- ++ (-0.4,0.) node[left, xshift=-0.3cm] {$\mu$};
\draw[thick] (0.2,6) -- ++ (-0.4,0.) node[left, xshift=-0.3cm] {$\Lambda$};
\draw[dashed, thick] (0.2,6) -- (6.,6);
\draw (0,2) node [right, xshift=1.0cm]{\bf{SM}};
\draw (0,5) node [right, xshift=0.5cm]{\bf{N-HDM}};

\draw[solid,thick, {Stealth[black]}-{Stealth[black]}, ] (3,4.1) -- (3,5.9);
\draw[solid,thick, {Stealth[black]}-{Stealth[black]}, ] (3,0.1) -- (3,3.9);

\draw (0,5) node [right, xshift=3.5cm]{$x^\alpha, \tilde{x}^\alpha, U$};

\draw (0,2) node [right, xshift=3.5cm]{$y^u, y^d, \bar{\theta}$};
\end{tikzpicture}

\caption{The framework: energy scales, EFTs and their flavor parameters. $\Lambda$ is the UV cutoff of theories with N Higgs doublets, all with the same gauge quantum numbers. All but one of these Higgs doublets acquires a mass of order M. Below M, the EFT is the minimal SM. \label{fig:EScheme}}
\end{figure}

In this theory, the Yukawa interactions are described by $ {\cal L}_Y = x_{ij}^\alpha \; q_i \bar{u}_j \phi_\alpha + \tilde{x}_{ij}^\alpha \; q_i \bar{d}_j \phi_\alpha^{\dagger}$, where $q_i$ are quark doublets and $\bar{d}_j, \bar{u}_j$ are singlet anti-quarks, all left-handed Weyl fields, and the Higgs doublets are $\phi_\alpha$, with $\alpha = 1... N$. These fields transform under a ``Higgs-Flavor" symmetry, $G_{\rm HF}$, so that the Yukawa couplings $x_{ij}^\alpha$, and $\tilde{x}_{ij}^\alpha$ are sparse with many zero entries. The Higgs fields all transform differently under $G_{\rm HF}$ so that the Higgs potential can be written in the form
\begin{align}
	\begin{split}
    V = &M^2_\alpha \phi_\alpha^\dagger \phi_\alpha + \mu^2_{\alpha \beta} \phi_\alpha^\dagger \phi_\beta
    + \lambda_{\alpha \beta}\, \phi_\alpha^\dagger \phi_\alpha \phi_\beta^\dagger \phi_\beta\\
    &+ \lambda_{\alpha \beta \gamma \delta}\,\phi_\alpha^\dagger \phi_\beta \phi_\gamma^\dagger \phi_\delta + h.c.
    \end{split}
    \label{eq:Higgspot}
\end{align}
While the quartics $\lambda_{\alpha\beta}$ are present in all models, the quartics $\lambda_{\alpha\beta\gamma\delta}$ are model-dependent, occurring only when there happens to be a $G_{\rm HF}$-invariant with at least three of $(\alpha,\beta,\gamma,\delta)$ different.  $CP$ and $G_{\rm HF}$ are conserved in all interactions and masses, except for the soft mass terms described by $\mu^2_{\alpha\beta}$, which are complex and have $\alpha \neq \beta$. While $|\mu^2_{\alpha\beta}|$ may take a variety of values, they are typified by a scale $\mu^2$. $N-1$ of the Higgs doublets have mass of order the scale $M$ and the other is fine-tuned to be much lighter. Hence the relevant mass scales and EFTs for this framework are illustrated in Fig.~\ref{fig:EScheme}. The unitary matrix U rotates the Higgs fields from flavor to mass eigenstates, $\phi_\alpha = U_{\alpha \beta} \, h_\beta$, and only $h_1$, which denotes the SM Higgs, acquires a vacuum expectation value, $\langle h_1 \rangle = v \simeq 174$ GeV. Below the scale $M$, we obtain the SM EFT with SM Yukawa coupling matrices
\begin{align}
    y^u_{ij} =  x^\alpha_{ij} \; U_{\alpha 1}\,, \qquad y^d_{ij} =  \tilde{x}^\alpha_{ij} \; U^{*}_{\alpha 1}.
    \label{eq:y}
\end{align}
The phases appearing in $y^u$ and $y^d$ arise from matrix elements of $U$ and $U^*$, and hence are correlated. We study theories where the Yukawa interactions have flavor determinants that involve a {\it single} product of scalar fields: $det(x^\alpha \phi_\alpha) \propto \Pi(\phi_\alpha)$ and $det(\tilde{x}^\alpha \phi^*_\alpha) \propto \Pi(\phi^*_\alpha)$. The determinants of $y^u$ and $y^d$ then involve the same single product of $U_{\alpha 1}$ matrix elements, 
\begin{align}
    det(y^u) \propto \Pi(U_{\alpha 1})\,, \hspace{0.5in} det(y^d) \propto \Pi(U^*_{\alpha 1})\,,
    \label{eq:strongCPsol}
\end{align}
so that $\bar{\theta}$ vanishes at tree-level.  For example, in the theory with three Higgses studied below, the Yukawa matrices are such that $\det(y^u) \propto U_{11} U_{31} U_{31}$ and $\det(y^d) \propto U_{11}^* U_{31}^* U_{31}^*$, even though the textures of the two matrices are quite different. However, the experimental limit from the nEDM is so severe that $\bar{\theta} = 0$ at tree-level may not be sufficient to ensure that these theories solve the strong CP problem.  Given the softness of CP violations, loop contributions to $\bar{\theta}$ arise as threshold corrections at the mass scale $M$ of the heavy Higgs doublets and take the form $\delta \bar{\theta} = f(x^\alpha_{ij}, \tilde{x}^\alpha_{ij}, U_{\alpha \beta})$.\\
In the SM there are two distinct small number problems associated with the quark mass matrices $y^{u,d}$: the strong CP problem and the flavor hierarchy problem. In the latter, small Yukawa couplings are needed to account for the small quark masses and mixings. From (\ref{eq:y}) we see that small values for some matrix elements of $y^{u,d}$ may be partially accounted for by small values of $|U_{\alpha1}|$ for $\alpha \neq 1$, that is by $\mu \ll M$. At the same time $\delta \bar{\theta}$ is proportional to $U_{\alpha \beta}$, so that it is advantageous for them to be small. Hence our framework offers the prospect of making progress on small flavor hierarchies as well as solving the strong CP problem.

\vspace{0.2cm}
{\bf Radiative Corrections.---}
Radiative corrections to Yukawa interactions modify the quark mass matrices and can lead to
a non-vanishing value for $\bar{\theta}= \arg\det(y^{u}y^{d})$. 
Below the scale $M$, the SM provides the effective description of the model, and it was shown by Ellis and Gaillard~\cite{Ellis:1978hq} that the first finite contribution to $\bar{\theta}$ appears at three-loop order (giving $\bar{\theta} \sim 10^{-16}$) and there is no divergent contribution before seven-loops. Between $\Lambda$ and $M$ there may be additional contributions through loop diagrams with a scalar exchange~\footnote{There are no contributions from wave function renormalization as they have the structure $\delta y^u = y^u B$ where $B$ is a hermitian matrix.}, illustrated in Fig.~\ref{fig:1loopDiagrams}.
\begin{figure}[tbh]
    \centering  
    \begin{tikzpicture}
    \begin{feynman}
        \vertex (L) at (-3,0);
        \vertex (R) at (+3,0);
        \vertex (M) at (0,0);
        \vertex (T) at (0,1.5);
        \vertex (LC) at (-1.5,0);
        \vertex (RC) at (+1.5,0);
        \vertex (B) at (0,-2) {$H_1$};
        \diagram*{
            (L) -- [fermion,edge label' = {$q_i$}] (LC),
            (LC) -- [anti fermion, edge label' = {$\bar d_l$}] (M),
            (M) -- [fermion, edge label' = {$q_k$}] (RC),
            (RC) -- [anti fermion, edge label' = {$\bar u_j$}] (R),
            (M) -- [anti charged scalar] (B),
            (LC) -- [charged scalar, half left, edge label={$H_\delta$}] (RC)
            };
    \end{feynman}
    \end{tikzpicture}
\caption{Feynman diagrams for the one-loop CP violating contributions to the quark Yukawa couplings $x_{ij}^\alpha$. Similar diagrams contribute to $\tilde{x}_{ij}^\alpha$.}
\label{fig:1loopDiagrams}
\end{figure}
\linebreak
The corrections to the SM Yukawa coupling matrix $y^{u}$ at renormalization point $\mu$ reads
\begin{align}
\begin{split}
\delta y^u = \frac{1}{16\pi^2}&\tilde{x}^\alpha \tilde{x}^{\dagger \beta}x^\gamma\;U_{\beta 1}\\ &\sum_{\delta} U^{*}_{\alpha \delta}  U_{\gamma \delta} \bigg( \frac{1}{\epsilon} + 1 + \log\bigg(\frac{\mu^2}{M_\delta^2}\bigg) \bigg)\,,
\label{eq:1loopRes}
\end{split}
\end{align}
and a similar expression for $\delta y^d$. Here dimensional regularization has been used and sums on $\alpha, \beta$ and $\gamma$ are understood.
The index $\delta$ runs over the $N$ mass eigenstate scalars, $h^\delta$.
As already commented above, the softness of CP violations ensures that loop contributions to $\bar{\theta}$ arise as threshold corrections at the mass scale $M$. In fact, the unitarity of the $U$ matrix implies that terms involving $(1/\epsilon + 1 + \log \mu^2)$ are proportional to $\tilde{x}^\alpha \tilde{x}^{\dagger \beta}x^\alpha\;U_{\beta 1}$ and $x^\alpha x^{\dagger \beta}\tilde{x}^\alpha\;U^{*}_{\beta 1}$ for $\delta y^u$ and $\delta y^d$, respectively. One can easily prove that $\tilde{x}^\alpha \tilde{x}^{\dagger \beta}x^\alpha\, (x^\alpha x^{\dagger \beta}\tilde{x}^\alpha)$ has the same matrix structure of $x^{\beta} (\tilde{x}^{\beta})$ and since at tree-level $y^u = x^\beta U_{\beta 1}$ and $y^d = \tilde{x}^\beta U^{*}_{\beta 1}$, this does not lead to corrections to $\bar{\theta}$ in the multi-Higgs theory. Keeping only the $\log(M_\delta^2)$ terms relevant for $\bar{\theta}$
\begin{align}
\begin{split}
\delta y^u = \frac{1}{16\pi^2}\tilde{x}^\alpha \tilde{x}^{\dagger \beta}x^\gamma\;&U_{\beta 1}  \bigg[U^{*}_{\alpha 1}  U_{\gamma 1}\log\left(\frac{M_2^2}{m_h^2}\right)\\
 & + \sum_{\delta=3}^N U^{*}_{\alpha \delta}  U_{\gamma \delta} \log\left(\frac{M_2^2}{M_\delta^2}\right)\bigg]\,,
\end{split}
\label{eq:1loopMain}
\end{align}
and a similar result follows for $\delta y^d$.
The first term above is the one-loop correction due to the SM Higgs in the SM EFT below $M_2$, which does not contribute to $\bar{\theta}$~\cite{Ellis:1978hq}. The only contribution to $\bar{\theta}$ comes from the second term in Eq.~\ref{eq:1loopMain} and depends logarithmically on the mass ratios of the heavy scalars.

An important result follows from Eq.~\ref{eq:1loopMain}: there are no corrections to $\bar{\theta}$ from one-loop scalar exchange in theories with two Higgs doublets. A similar statement can be made for higher order corrections to the quark Yukawa matrices. Below M we recover the SM EFT and therefore there is a negligible finite contribution to $\bar{\theta}$ at 3-loops and no divergent contributions up to 6 loops~\cite{Ellis:1978hq}. Above M, where the two Higgs doublet model is the correct EFT, the finite corrections to the Yukawa matrices can be: constant terms, terms proportional to powers of $m_h/M_2$ and terms proportional to $\log(M_2/m_h)$ (due to the
softness of CP violation). At 2-loops we checked diagram by diagram that the constant pieces do not contribute to $\bar{\theta}$. This is easy to prove using the unitarity of the U matrix and noting that each propagator for the scalar $H_{\beta}$ comes with the insertion of a $U_{\alpha\beta}$ and a $U^{*}_{\gamma\beta}$ matrix. Terms proportional to powers of $m_h/M_2$ correspond to higher dimensional operators in the SM EFT and are negligible in the limit where $M_2 \gg m_h$. Finally, terms proportional to $\log(M_2/m_h)$ can be understood as coming from RG scaling in the SM EFT below the heavy Higgs mass, where we know there are no corrections to $\bar{\theta}$ at two loops. Similar arguments apply to any loop order; in theories with two Higgs doublets, we conclude there are no radiative corrections to $\bar{\theta}$ beyond those of the SM. On the other hand, for theories with $N > 2$, there are corrections to $\bar{\theta}$ already at one loop. These are proportional to a combination of the $x, \tilde{x}$ and $U$ matrices and are therefore model-dependent. In general, one needs to compute $\bar{\theta} = \arg\det[(y^u + \delta y^u)(y^d + \delta y^d)]$ to assess whether a theory solves the strong CP problem for a reasonable choice of parameters.   

\vspace{0.2cm}
{\bf Models.---}
A complete study and categorization of theories within the framework described above is beyond the scope of this work. In the following, we  discuss a representative example that captures the main features of this class of models.\\

We consider as an example theories with 3 Higgs doublets and $G_{\rm HF}=\rm U(1)$. Restricting the charges of scalar and quark fields to be smaller than two, and requiring the theory to generate quark masses, CKM mixing angles at first order in the off-diagonal Yukawa entries, and a non-zero Jarkslog invariant, we found $2160$ possible charge assignments. In each of these theories the scalar sector is the same, with three real diagonal and three complex off-diagonal mass parameters. For simplicity and with the same conventions as in Eq.~\ref{eq:Higgspot}, we assume $\epsilon_{\alpha \beta} = \mu_{\alpha \beta}^2/M^2 \ll 1$, such that the off-diagonal elements of the rotation matrix are $U_{ij} = -U_{ij}^* \simeq \epsilon_{ij}e^{i\beta_{ij}}$, while $U_{ii} \simeq 1$. The number of parameters in the quark Yukawas matrices depends on the number of non-zero entries as dictated by the flavor symmetry. A representative $\rm U(1)_{\rm HF}$ charge assignment, which has the virtue of reducing the number of free parameters, is shown in Tab.~\ref{tab:charges3H_2}, leading to the Yukawa interactions
\begin{table}[tb]
    \centering
    \begin{tabular}{c|c c c | c c c | c c c | c c c}
            & $\phi_1$ & $\phi_2$ & $\phi_3$ & $q_1$ & $q_2$ & $q_3$ & $\bar{u}_1$ & $\bar{u}_2$ & $\bar{u}_3$ & $\bar{d}_1$ & $\bar{d}_2$ & $\bar{d}_3$\\
            \hline
       ${\rm U(1)}_{\rm \rm HF}$  & 0 & 2 & -1 & 0 & 1 & -1 & 2 & -1 & 1 & -2 & -1 & 1
    \end{tabular}
    \caption{$\rm U(1)_{\rm HF}$ charges for the representative 3 Higgs doublet model discussed in this letter.}
    \label{tab:charges3H_2}
\end{table}
\begin{align}
\begin{split}
     \mathcal{L}_{\rm Y} =&
     \begin{pmatrix}
         q_1 & q_2 & q_3
     \end{pmatrix}\begin{pmatrix}
        0 & 0 & x_{13}\phi_3\\
        0 & x_{22}\phi_1 & 0 \\
        x_{31}\phi_3 & x_{32}\phi_2 & x_{33}\phi_1 
    \end{pmatrix}\begin{pmatrix}
        \bar u_1 \\ \bar u_2 \\ \bar u_3
    \end{pmatrix}\\
    +& \begin{pmatrix}
         q_1 & q_2 & q_3
     \end{pmatrix}\begin{pmatrix}
        0 & \tilde{x}_{12}\phi_3^\dagger & 0 \\
        \tilde{x}_{21}\phi_3^\dagger & \tilde{x}_{22}\phi_1^\dagger & \tilde{x}_{23}\phi_2^\dagger \\
        0 & 0 & \tilde{x}_{33}\phi_1^\dagger
    \end{pmatrix}\begin{pmatrix}
        \bar d_1 \\ \bar d_2 \\ \bar d_3
    \end{pmatrix}\,.
\end{split}
\label{eq:textureModelII}
\end{align}
The SM Yukawa coupling matrices are obtained by inserting $\phi_\alpha = U_{\alpha 1} h_1$. At tree-level, $\bar{\theta} = 0$ as $\det(y^u) = x_{13}x_{22}x_{31}U_{11}U_{31}U_{31}$ and $ \det(y^d) = \tilde{x}_{12}\tilde{x}_{21} \tilde{x}_{33}U^*_{11}U^*_{31}U^*_{31}$.
This theory has 13 real parameters and 3 phases. Quark masses and mixings fix 9 real parameters and the CKM fixes one phase, according to
\begin{align}
\begin{split}
    &y_u \simeq x^u_{31}|U^*_{31}V_{ub}|\,,\quad y_c \simeq x^u_{22}|U_{11}|\,,\quad
    y_t \simeq x^u_{33}|U_{11}|\,,\\
    &y_d \simeq x^d_{21}|U^*_{31}V_{us}|\,,\quad
    y_s \simeq x^d_{22}|U^*_{11}|\,,\quad
    y_b \simeq x^d_{33}|U^*_{11}|\,,\\
    &|V_{us}| \, y_s \simeq  x^d_{12}|U^*_{31}| \,,\quad 
    |V_{ub}| \, y_t \simeq x^u_{13}|U_{31}|\,,\\
    &|V_{cb}| \, y_b \simeq  x^d_{23}|U^*_{21}|\,,\quad \delta_{\rm CKM} \simeq \beta_{12}+2\beta_{13}\,,
\end{split}
\label{eq:CKMconstraintsII}
\end{align}
to leading order for each of $V_{ub}$ (from $y^u$), $V_{us}$ and $V_{cb}$ (from $y^d$). This leaves four free real parameters, $|U_{21}|,|U_{31}|,|U_{23}|,x_{32}$.
Requiring $x_{13}$ and $\tilde{x}_{23}$ to be perturbative, gives the lower bounds $\ep{13}\gtrsim 0.004$, and $\ep{12}\gtrsim 0.001$. 
Furthermore, we have the requirement of sufficiently small one-loop corrections to $\bar{\theta}$. Using Eq.~\ref{eq:1loopMain},  setting $\epsilon_{23}$ to its minimal natural value, $\epsilon_{23} \simeq \epsilon_{12} \epsilon_{13}$, and taking phases of order unity, the dominant terms are
\begin{align}
      \delta\bar{\theta} \simeq 
      10^{-5} x_{32} \, \epsilon_{12} - 3 \times 10^{-9} \epsilon_{13}^2.    \label{eq:thetabarII2}
\end{align}
For example, taking $\epsilon_{12} \simeq 10^{-3}, \, \epsilon_{13} \simeq 10^{-2}, \, \epsilon_{23} \simeq 10^{-5}$
and $x_{32} \simeq 10^{-2}$ gives $\delta\bar{\theta} \simeq 10^{-10}$, with numerical values of the Yukawa matrices
\begin{align}
\begin{split}
     x \simeq 
     \begin{pmatrix}
        0 & 0 & 0.4\\
        0 & 0.005 & 0 \\
        0.1 & 0.01 & 1 
    \end{pmatrix},\;
    \tilde{x} \simeq
    \begin{pmatrix}
        0 & 0.006 & 0 \\
        0.005 & 0.0003 & 0.8  \\
        0 & 0 & 0.02 
    \end{pmatrix}.
\end{split}
    \label{eq:xII}
\end{align}
This theory illustrates how $G_{\rm HF}$ leads to a close relation between the strong CP and flavor problems.  In the SM all we know about the quark Yukawa couplings $y^{u,d}$ is that they are general complex matrices that have certain eigenvalues, ranging over 5 orders of magnitude, and are diagonalised by certain mixing angles. In this theory, solving the strong CP problem leads to a very different set of real 
flavor matrices shown in (\ref{eq:xII}). Many entries vanish, and the hierarchies in $x$ are much less than the hierarchies in the up-type quark masses. Of course, we do not provide an understanding for these flavor  matrices, but stress that the flavor problem has been transformed by our attempt to solve the strong CP problem. In the supplemental material we discuss theories with $N=2$ and a different theory with N=3, highlighting the vast and diverse space of models within in our framework.

\vspace{0.2cm}
{\bf Conclusion.---}
Building on early work \cite{Georgi:1978xz, Mohapatra:1979kg, Glashow:2001yz}, we introduced a new framework for studying flavor and the strong CP problem. The scalar sector of the SM is augmented from $1$ to $N$ Higgs doublets, and there are no additions to the fermion or gauge sectors. All dimension 4 interactions are symmetric under $CP$ and a Higgs-Flavor symmetry $G_{\rm HF}$ carried by both quarks and Higgs doublets. The SM Yukawa matrices are replaced by real couplings, due to $CP$, and have many zero entries, due to $G_{\rm HF}$. $CP$ and $G_{\rm HF}$ are softly broken via the off-diagonal entries of the Higgs mass matrix. We restricted our attention to theories that have $N-1$ heavy Higgs doublets with mass of order $M$ and one much  lighter Higgs $h$.  Thus the SM appears as the EFT below M, with Yukawa coupings and strong CP parameter arising from integrating out the heavy Higgs doublets. In this framework, the flavor and strong CP problems are inextricably linked.

We have computed the 1-loop threshold corrections to $\bar{\theta}$ resulting from integrating out the heavy Higgs doublets at the mass scale $M$. In this class of theories, all non-vanishing contributions are proportional to $\log(M_\alpha/M_\beta)$, where $\alpha, \beta$ label the heavy Higgs doublet mass eigenstates. Thus $N=2$ theories, with a single heavy Higgs, don't have any contribution to $\bar{\theta}$ at 1-loop. We studied radiative contributions to $y^u$ and $y^d$ at two loop order with $N=2$ and find only contributions from RG scaling in the SM EFT below $M$, which are known to not induce $\bar{\theta}$. Hence models with $N=2$ are particularly powerful at solving the strong CP problem. Contributions from higher-dimensional operators at the scale $\Lambda$ in general depend on the UV completion of these multi-Higgs theories but they are proportional to $(M^2/\Lambda^2) \ln (\Lambda^2/M^2)$. Order of magnitude estimates, allow us to conclude that they are negligible if the scale $M$ is a few orders of magnitude below $\Lambda$. Furthermore, contributions to the neutron EDM through higher dimensional operators at the scale $M$ are negligible if the latter is taken to be well above the electroweak scale.  

In this framework, the requirement of a ``Higgs-Flavor" symmetry that gives $\bar\theta=0$ at tree-level may be a useful guide in constructing predictive theories of flavor. A small value for $\mu/M$ may suppress radiative corrections to $\bar{\theta}$ and contribute to the smallness of quark masses and mixing angles. 

This framework is easily extended to the lepton sector to give a fully realistic description of nature, whether neutrinos are Majorana or Dirac. This leads to relations between the phases appearing in the quark and lepton mass matrices with important implications for models where the cosmological baryon asymmetry arises from leptogenesis. Moreover, $G_{\rm HF}$ charges can be found for quarks and leptons that are consistent with either $\rm SU(5)$ or $\rm SO(10)$ unification. Finally, our framework offers diverse and model-dependent experimental signatures, including flavor observables like K, B, and D mixing, as well as opportunities for direct searches at current and future collider experiments. Given the richness of the phenomenology in this class of multi-Higgs theories, we have intentionally remained agnostic on the UV completion and we defer its discussion to future works.

\vspace{1cm}

\begin{centering}
\textbf{\textit{Acknowledgments}}\\

We thank Quentin Bonnefoy for useful discussions. This work is supported by the Office of High Energy Physics of the U.S. Department of Energy under contract DE-AC02-05CH11231 and by the NSF grant PHY-2210390.
\end{centering}

%%%%%%%%%%%%%%%%%%%%%%%%%%%%%%%%%%%%%%%
%%%%%%%%%%%%%%%%%%%%%%%%%%%%%%%%%%%%%%%
%\clearpage
\onecolumngrid
\appendix

\section{Small Parameters and $U(1)^9$}\label{app:U19}

In the class of theories discussed in this letter, although many entries in the Yukawa matrices $x^\alpha_{ij}, \tilde{x}^\alpha_{ij}$ vanish, the non-zero entries may not all be determined from data on quark masses and mixings. Furthermore, radiative corrections to $\bar{\theta}$ depend on these unknown parameters. While $\bar{\theta} < 10^{-10}$ results if these parameters are sufficiently small, how can we determine if such small values constitute a solution to the strong CP problem?\\ 
One possibility is that the sizes of the $x$ and $\tilde{x}$ matrix elements are controlled by conventional flavor symmetries. A convenient and simple description may be possible in terms of symmetry breaking parameters $\epsilon_{q_i}, \epsilon_{u_i}, \epsilon_{d_i}$ of the flavor symmetry $\rm U(1)^3_q \times U(1)^3_u \times U(1)^3_d$ such that, when non-zero, $x^\alpha_{ij} \sim \epsilon_{q_i} \epsilon_{u_i}$ and  $\tilde{x}^\alpha_{ij} \sim \epsilon_{q_i} \epsilon_{d_i}$. Where possible, in computing the value of radiative corrections to $\bar{\theta}$, we estimate any 
$x^\alpha_{ij}, \tilde{x}^\alpha_{ij}$ that cannot be determined from data using this approximate flavor symmetry. The 9 real symmetry breaking parameters $\epsilon_{q_i}, \epsilon_{u_i}, \epsilon_{d_i}$ are all $\lesssim 1$ and can be determined from the 6 quark masses and 3 quark mixing angles, so that $\bar{\theta}$ can be estimated in terms of $U_{\alpha \beta}$. In fact, since the top Yukawa is close to unity it determines two of these parameters, $\epsilon_{q_3} \sim \epsilon_{u_3} \sim 1$, implying that fitting to quark masses and mixings leads to a relation between the various $U_{\alpha1}$.  Using this procedure, the radiative correction to $\bar{\theta}$ can then be evaluated in terms of the remaining independent values of $U_{\alpha \beta}$, allowing a determination of whether the strong CP problem is naturally solved. The size of the required $\epsilon_{q_i}, \epsilon_{u_i}, \epsilon_{d_i}$ parameters will also allow an assessment of any progress on the flavor problem. On the other hand, for some models a fit to the quark masses and mixings using natural values for $U_{\alpha \beta}$ yields values of $x^\alpha_{ij}, \tilde{x}^\alpha_{ij}$ that cannot be consistently understood from an approximate $\rm U(1)^9$ flavor symmetry (see the theory discussed in the main text). In this case, one can derive upper bounds on those $x, \tilde{x}$ matrix elements not determined from data in order that $\bar{\theta} < 10^{-10}$.  For these models, an assessment of whether the strong CP problem is solved can be made according to how small these parameters must be made.

\section{Theories with N=2 and $G_{\rm \rm HF} = \mathbb{Z}_3$}\label{app:N2models}

We found 864 models with two Higgs bosons and $G_{\rm HF}=\mathbb{Z}_3$ that could accommodate quark masses, CKM angles (from diagonalization of Yukawa matrices at first order), a nonzero Jarlskog invariant and $\bar\theta=0$ at tree level. Notably, all of these theories have precisely three zeros in each Yukawa coupling matrix at tree level. Among these theories there occur 24 unique textures of Yukawa matrices that appear in 48 unique $\{y^u,y^d\}$ pairings. Here by ``texture" we mean the full form of the tree-level Yukawas, keeping track of where the $U_{21}$ and $U_{11}$ factors appear as well as zeros. Each of the 24 unique textures of $y^u$ corresponds to one of the 24 textures of $y^d$ by complex conjugation, as one would expect at the level of charges. As such we list the 24 textures in the following format: $0$ where a coupling is absent at tree level, $1$ where the coupling to $\phi_1$ (or $\phi_1^*$) is allowed and thus a factor of $U_{11}$ is present, and $\epsilon$ where the coupling to $\phi_2$ (or $\phi_2^*$) is allowed and thus a factor of $U_{21}$ (or $U_{21}^*$) is present. For example, the texture corresponding to the theory discussed in the next section would be written as:
\begin{align}
    y^u \sim& \begin{pmatrix}
        1 & 0 & 0 \\
        0 & \epsilon & \epsilon \\
        \epsilon & 1 & 1
    \end{pmatrix}\,,
    &
    y^d \sim& \begin{pmatrix}
        1 & 1 & \epsilon \\
        \epsilon & \epsilon & 0 \\
        0 & 0 & 1
    \end{pmatrix}\,.
\end{align}
With this convention, the 24 textures are listed in Tab.~\ref{tab:N2Textures}. They are arbitrarily numbered for reference, e.g. the above theory would be said to have texture $\{3,14\}$. Note that for each of these matrices there is another one obtained exchaning $\epsilon \leftrightarrow 1$. The 24 distinct pairing of textures are listed in Tab.~\ref{tab:Z3Pairings}. 
\begin{table}
\centering
\begin{align}\nonumber
\begin{array}{cccccc}
 \text{1:} \left(
\begin{array}{ccc}
 1 & 1  & 0  \\
 0 & 0 & \epsilon \\
 \epsilon & \epsilon & 1 \\
\end{array}
\right) & \text{2:} \left(
\begin{array}{ccc}
 1  & 0 & 1  \\
 0 & \epsilon  & 0 \\
 \epsilon & 1 & \epsilon \\
\end{array}
\right) & \text{3:} \left(
\begin{array}{ccc}
 1 & 0 & 0 \\
 0  & \epsilon & \epsilon \\
 \epsilon & 1  & 1  \\
\end{array}
\right) & \text{4:} \left(
\begin{array}{ccc}
 1 & \epsilon & 1 \\
 0 & 1  & 0 \\
 \epsilon  & 0 & \epsilon  \\
\end{array}
\right) & \text{5:} \left(
\begin{array}{ccc}
 0  & 1  & 1 \\
 \epsilon & 0 & 0 \\
 1 & \epsilon & \epsilon  \\
\end{array}
\right) & \text{6:} \left(
\begin{array}{ccc}
 0 & 1 & 0 \\
 \epsilon  & 0  & \epsilon \\
 1 & \epsilon & 1  \\
\end{array}
\right) \\[0.5cm]
 \text{7:} \left(
\begin{array}{ccc}
 0 & 0 & 1  \\
 \epsilon  & \epsilon  & 0 \\
 1 & 1 & \epsilon \\
\end{array}
\right) & \text{8:} \left(
\begin{array}{ccc}
 0 & \epsilon  & 0 \\
 \epsilon & 1 & \epsilon \\
 1  & 0 & 1  \\
\end{array}
\right) & \text{9:} \left(
\begin{array}{ccc}
 \epsilon  & 1 & 1 \\
 1 & 0 & 0 \\
 0 & \epsilon  & \epsilon  \\
\end{array}
\right) & \text{10:} \left(
\begin{array}{ccc}
 \epsilon & 0  & 0  \\
 1 & \epsilon & \epsilon \\
 0  & 1 & 1 \\
\end{array}
\right) & \text{11:} \left(
\begin{array}{ccc}
 \epsilon  & \epsilon & 1  \\
 1 & 1 & 0 \\
 0 & 0  & \epsilon \\
\end{array}
\right) & \text{12:} \left(
\begin{array}{ccc}
 \epsilon & \epsilon & 0  \\
 1 & 1 & \epsilon \\
 0  & 0  & 1 \\
\end{array}
\right) \\[0.5cm]
 \text{13:} \left(
\begin{array}{ccc}
 1 & 1 & 0 \\
 \epsilon & \epsilon  & 1  \\
 0  & 0 & \epsilon \\
\end{array}
\right) & \text{14:} \left(
\begin{array}{ccc}
 1 & 1 & \epsilon \\
 \epsilon  & \epsilon & 0  \\
 0 & 0 & 1 \\
\end{array}
\right) & \text{15:} \left(
\begin{array}{ccc}
 1  & 0  & 0 \\
 \epsilon & 1 & 1  \\
 0 & \epsilon & \epsilon \\
\end{array}
\right) & \text{16:} \left(
\begin{array}{ccc}
 1 & \epsilon & \epsilon \\
 \epsilon & 0  & 0 \\
 0  & 1 & 1  \\
\end{array}
\right) & \text{17:} \left(
\begin{array}{ccc}
 0 & 1 & 0 \\
 1  & \epsilon & 1 \\
 \epsilon & 0  & \epsilon  \\
\end{array}
\right) & \text{18:} \left(
\begin{array}{ccc}
 0  & 0  & \epsilon \\
 1 & 1 & 0 \\
 \epsilon & \epsilon & 1 \\
\end{array}
\right) \\[0.5cm]
 \text{19:} \left(
\begin{array}{ccc}
 0 & \epsilon & 0 \\
 1 & 0 & 1  \\
 \epsilon  & 1  & \epsilon \\
\end{array}
\right) & \text{20:} \left(
\begin{array}{ccc}
 0 & \epsilon  & \epsilon \\
 1 & 0 & 0  \\
 \epsilon & 1 & 1 \\
\end{array}
\right) & \text{21:} \left(
\begin{array}{ccc}
 \epsilon  & 1 & \epsilon \\
 0 & \epsilon  & 0  \\
 1 & 0 & 1 \\
\end{array}
\right) & \text{22:} \left(
\begin{array}{ccc}
 \epsilon & 0  & 0 \\
 0 & 1 & 1 \\
 1  & \epsilon & \epsilon  \\
\end{array}
\right) & \text{23:} \left(
\begin{array}{ccc}
 \epsilon  & 0 & \epsilon \\
 0 & 1 & 0 \\
 1 & \epsilon  & 1 \\
\end{array}
\right) & \text{24:} \left(
\begin{array}{ccc}
 \epsilon & \epsilon & 0 \\
 0  & 0  & 1 \\
 1 & 1 & \epsilon  \\
\end{array}
\right) \\
\end{array}
\end{align}
\caption{The 24 unique Yukawa textures that appear in viable $N=2$ theories with $G_{\rm HF}=\mathbb{Z}_3$.}
\label{tab:N2Textures}
\end{table}

\begin{table}[H]
    \centering
\begin{tabular}{c|c}
$y^u$ & $y^d$ \\\hline
 1 & 16,21 \\
 2 & 13 \\
 3 & 14 \\
 4 & 15,17,24 \\
 5 & 13 \\
 6 & 14 \\
 7 & 17 \\
 8 & 16,18,21 \\
 9 & 15,17,24 \\
 10 & 16,21 \\
 11 & 13,19,22 \\
 12 & 14,20,23 \\
\end{tabular}
    \caption{All 24 pairs of Yukawa textures in viable $G_{HF}=\mathbb{Z}_3$ models. Textures are labeled according to Tab.~\ref{tab:N2Textures}.}
    \label{tab:Z3Pairings}
\end{table}

\section{A Theory with N=2}\label{N=2Theory}

Theories with two Higgs doublets are particularly simple and interesting. CP and Higgs flavor symmetry is broken by a single 
soft scalar mass $\mu_{12}^2 \equiv \mu^2 e^{i\beta}$ (with $\mu^2\,, \beta \in \mathbb{R}$). The unitary matrix transforming from Higgs flavor to mass eigenstates is
\begin{align}
	U = \begin{pmatrix}
		\cos\xi & -\sin\xi\, e^{i\beta}\\
		\sin\xi\, e^{-i\beta} & \cos\xi
	\end{pmatrix}\,,
\end{align}
with $(\cos2\xi, \, \sin2\xi) = \frac{(M_1^2-M_2^2, \, 2 \mu^2)}{\sqrt{(M_1^2-M_2^2)^2+4\mu^4}}$.
Here we study the case of a $\mathbb{Z}_3$ Higgs flavor symmetry, yielding ${\rm{det}}(y^uy^d) \in \mathbb{R}$. A discussion of these theories and realistic charge assignments can be found in the supplemental material. As discussed in the main text, there are no radiative corrections to $\bar{\theta}$ in addition to those already present in the SM and therefore the strong CP problem is solved in these theories.
A two Higgs doublet theory at the TeV scale with a $\mathbb{Z}_3$ symmetry was discussed by Ref.~\cite{Ferreira:2019aps}. After electroweak symmetry breaking, they found that CP violation appeared only in the $W^\pm$ and $H^\pm$ couplings, and that the 1-loop contribution to $\bar{\theta}$ from a charged Higgs loop vanished, in agreement with our more general discussion.
\begin{table}[h!]
	\centering
	\begin{tabular}{c|c c| c c c | c c c | c c c}
		& $\phi_1$  & $\phi_2$ & $q_1$ & $q_2$ & $q_3$ & $\bar{u}_1$ & $\bar{u}_2$ & $\bar{u}_3$ & $\bar{d}_1$ & $\bar{d}_2$ & $\bar{d}_3$ \\
		\hline $\mathbb{Z}_3$ & 1 & $\alpha^2$ & $\alpha^2$ & $\alpha$ & 1 & $\alpha$ & 1 & 1 & $\alpha$ & $\alpha$ & 1
	\end{tabular}
	\caption{The Higgs Flavor charges in a two Higgs doublet model with $G_{\rm HF}$ taken to be a discrete $\mathbb{Z}_3$ global symmetry. Here $\alpha$ is the generator of $\mathbb{Z}_3$ -- i.e. $\alpha^3=1$. %Note that $\alpha^* = \alpha^2$.
	}
	\label{tab:Z3theory}
\end{table}
For concreteness we discuss the model with charges given in Table~\ref{tab:Z3theory} leading to Yukawa interactions 
\begin{align}
	\begin{split}
		\mathcal{L}_{\rm Y} &=
		\begin{pmatrix}
			q_1 & q_2 & q_3
		\end{pmatrix}\begin{pmatrix}
			x_{11}\phi_1 & 0 & 0 \\
			0 & x_{22}\phi_2 & x_{23}\phi_2 \\
			x_{31}\phi_2 & x_{32}\phi_1 & x_{33}\phi_1 
		\end{pmatrix}\begin{pmatrix}
			\bar u_1 \\ \bar u_2 \\ \bar u_3
		\end{pmatrix}\\
		&+ \begin{pmatrix}
			q_1 & q_2 & q_3
		\end{pmatrix}\begin{pmatrix}
			\tilde{x}_{11}\phi_1^\dagger & \tilde{x}_{12}\phi_1^\dagger & \tilde{x}_{13}\phi_2^\dagger \\
			\tilde{x}_{21}\phi_2^\dagger & \tilde{x}_{22}\phi_2^\dagger & 0 \\
			0 & 0 & \tilde{x}_{33}\phi_1^\dagger
		\end{pmatrix}\begin{pmatrix}
			\bar d_1 \\ \bar d_2 \\ \bar d_3
		\end{pmatrix}\,,
	\end{split}
	\label{eq:Yukawa2HDM}
\end{align}
where we suppress the scalar index on $x$ and $\tilde{x}$ as each entry of these matrices receives a contribution from only one scalar. 
Thus, in this model there are 13 real parameters (12 couplings $x_{ij}$, $\tilde{x}_{ij}$ and $\xi$) and one phase ($\beta$). The measured quark Yukawa eigenvalues, $y_i$, and CKM parameters constrain 10 of them.
Moreover, it is straightforward to check that $\det(y^u y^d)=0$ and therefore $\bar\theta=0$ at tree level. For the case that $\xi \ll 1$, the $x$ and $\tilde{x}$ matrices take the form
\begin{align}
	\begin{split}
		x &=  
		\begin{pmatrix}
			y_u & 0 & 0\\
			0 & y_c/\xi & V_{cb} \, y_t/ \xi \\
			x_{31} & x_{32} & y_t 
		\end{pmatrix}, \hspace{0.05in}
		\tilde{x} =
		\begin{pmatrix}
			y_d & V_{us} \, y_s & V_{ub} \, y_b / \xi\\
			\tilde{x}_{21} & y_s/ \xi & 0 \\
			0 & 0 & y_b 
		\end{pmatrix}.
	\end{split}
	\label{eq:xN=2}
\end{align}
The 11 and 33 entries of the $x, \tilde{x}$ matrices are the same as in the $y^u, y^d$ matrices, so there is no progress here on the flavor problem.  However, the 22 entries, and the entries generating CKM mixing can be much larger than in the SM Yukawas. For example all entries in the heavy $2 \times 2$ space of $x$ could be large and order unity. Note also that the requirement of zero $\bar{\theta}$ at tree level has forced 6 texture zeros.

\section{A Different N=3 Model}\label{app:N=3_2}

We study here another theory with N=3 and $G_{\rm HF} = \rm U(1)$ to show how a different charge assignment under the Higgs-flavor symmetry leads to a different result for flavor and the strong CP problem. The $\rm U(1)_{\rm HF}$ charges are as presented in Ref.~\cite{Glashow:2001yz}, and reported in Tab.~\ref{tab:chargesN3_2}.
\begin{table}[H]
    \centering
    \begin{tabular}{c|c c c| c c c| c c c| c c c}
            & $\phi_1$ & $\phi_2$ & $\phi_3$ & $q_1$ & $\bar{u}_1$ & $\bar{d}_1$ & $q_2$ & $\bar{u}_2$ & $\bar{d}_2$ & $q_3$ & $\bar{u}_3$ & $\bar{d}_3$\\
            \hline
       $\rm U(1)_{\rm HF}$  & 0 & 1 & 2 & 1 & -1 & -1 & 0 & 0 & 0 & -1 & 1 & 1
    \end{tabular}
    \caption{$\rm U(1)_{\rm F}$ charges for the theory with N=3 and $G_{\rm \rm HF} = \rm U(1)$ discussed in App.~\ref{app:N=3_2}}
    \label{tab:chargesN3_2}
\end{table}
The allowed Yukawa interactions are:
\begin{align}
     \mathcal{L}_{\rm Y} =&
     \begin{pmatrix}
         q_1 & q_2 & q_3
     \end{pmatrix}\begin{pmatrix}
        x_{11}\phi_1 & 0 & 0 \\
        x_{21}\phi_2 & x_{22}\phi_1 & 0 \\
        x_{31}\phi_3 & x_{32}\phi_2 & x_{33}\phi_1 
    \end{pmatrix}\begin{pmatrix}
        \bar u_1 \\ \bar u_2 \\ \bar u_3
    \end{pmatrix}
    + \begin{pmatrix}
         q_1 & q_2 & q_3
     \end{pmatrix}\begin{pmatrix}
        \tilde{x}_{11}\phi_1^\dagger & \tilde{x}_{12}\phi_2^\dagger & \tilde{x}_{13}\phi_3^\dagger \\
        0 & \tilde{x}_{22}\phi_1^\dagger & \tilde{x}_{23}\phi_2^\dagger \\
        0 & 0 & \tilde{x}_{33}\phi_1^\dagger
    \end{pmatrix}\begin{pmatrix}
        \bar d_1 \\ \bar d_2 \\ \bar d_3
    \end{pmatrix}\,.
    \label{eq:textureModelI}
\end{align}
This theory has 15 real parameters and 3 phases, and the pattern of quark masses and CKM parameters is easily reproduced. We obtain
\begin{align}
\begin{split}
    y_u &\simeq x_{11}|U_{11}|\,,\quad 
    y_c \simeq x_{22}|U_{11}|\,,\quad
    y_t \simeq x_{33}|U_{11}|\,,\\
    y_d &\simeq\tilde{x}_{11}|U^*_{11}|\,,\quad
    y_s \simeq \tilde{x}_{22}|U^*_{11}|\,,\quad
    y_b \simeq \tilde{x}_{33}|U^*_{11}|\,,\\
     |V_{us}| \, y_s &\simeq  \tilde{x}_{12}|U^*_{21}| \,,\quad |V_{ub}| \, y_b \simeq \tilde{x}_{13}|U^*_{31}|\,,\quad
    |V_{cb}| \, y_b \simeq  \tilde{x}_{23}|U^*_{21}|\,,\quad  \delta_{\rm CKM} \simeq 2\beta_{12}-\beta_{13}\,.
\end{split}
\label{eq:CKMconstraintsI}
\end{align}
Together with the requirement of perturbative Yukawa couplings, these results give the lower bounds $|U_{21}| \gtrsim 10^{-3}$ and $|U_{31}| \gtrsim 10^{-4}$. These relations and the unitarity of the U matrix leave only six real free parameters in this theory: $|U_{12}|, |U_{13}|, |U_{23}|, x^u_{21}, x^u_{31}, x^u_{32}$ and two phases. Using the result of Eq.~\ref{eq:1loopMain}, we compute the one-loop correction to $\bar{\theta}$, which in the small mixing approximation reads:
\begin{align}
\begin{split}
    \delta\bar{\theta} \simeq&\; \bigg(5\times 10^{-5}\frac{\ep{23}x^u_{21}}{\ep{31}}+6\times 10^{-10}\frac{\ep{23}x^u_{31}}{\ep{21}}-2\times 10^{-4}\ep{23}x^u_{21}x^u_{31} - 0.2\frac{\ep{21}\ep{23}x^u_{21}x^u_{32}}{\ep{31}} 
    \\
    &+ 4\times 10^{-4}\ep{21}\ep{23}\ep{31}x^u_{21}x^u_{32}\bigg)\sin(\beta_{21}+\beta_{23})\\
    & + \bigg( 5\times10^{-5} \ep{21}x^u_{21} + 6\times10^{-10}\ep{31}x^u_{31} - 3\times10^{-4}\ep{21}\ep{31}x^u_{21}x^u_{31}\bigg)\sin(2\beta_{21}-\beta_{31})\\
    & - \bigg( 6\times10^{-9}\frac{\ep{23}\ep{31}}{\ep{21}} - 3\times10^{-5}\ep{23}\ep{31}x^u_{21} - 2\times 10^{-5}\ep{23}\ep{31}x^u_{32}\bigg)\sin(\beta_{21}-\beta_{23}-\beta_{31}) \,.
    \label{eq,thetabarI}
\end{split}
\end{align}
To evaluate whether this radiative contribution is sufficiently small for the model to solve the strong CP problem, we are faced with the problem that $x_{21}, x_{31}$ and $x_{32}$ are not determined by data. Here we use an approximate $\rm U(1)^9$ flavor symmetry, described in App. \ref{app:U19}, and estimate 
\begin{align}
 x_{21} \simeq 2 \times 10^{-5} \, U_{21}\,, \hspace{0.3in} x_{31} \simeq 5 \times 10^{-4} \, U_{21}^2\,, \hspace{0.3in} x_{32}\simeq 10^{-1} \, U_{21}\,, \hspace{0.3in} U_{31} \simeq 0.5 \, U_{21}^2\,.
\end{align}
Furthermore, for the symmetry breaking parameter $\epsilon_{q_1}$ to be less than unity requires $U_{21} \gtrsim 0.1$. While $U_{32}$ is not determined by this approximate symmetry, it has the same Higgs-flavor charge as $U_{21}$, and hence we expect $U_{32} \simeq U_{21}$. Using these results, and taking $U_{21}$ near its lower bound, the first term of  (\ref{eq,thetabarI}) is of order $10^{-9}$, as estimated in \cite{Glashow:2001yz}, which within uncertainties is consistent with data. However the fourth term in (\ref{eq,thetabarI}), not considered in \cite{Glashow:2001yz}, is larger and is similarly estimated to be $10^{-8}$. Hence, we conclude that while this model makes significant progress on the strong CP problem, it does not convincingly solve it. Perhaps an alternative flavor symmetry scheme can be found that further suppresses these 1-loop corrections.

\section{Higher-Dimensional Operators}\label{sec:HDO}

Higher-dimensional operators, involving fields $q_i, \bar{u}_i, \bar{d}_i$ and $\phi_\alpha$, that can potentially spoil the solution to the strong CP problem, may be generated at the scale $\Lambda$ where the multi-Higgs theory is UV completed. The leading such operators occur at dimension 6
\begin{align}
	\begin{split}
		{\cal L}_6 \supset  \frac{C^{\alpha\beta\gamma}_{ij}}{\Lambda^2} \; q_i\bar u_j \, \phi_\alpha \phi_\beta \phi_\gamma^{\dagger} +
		\frac{\tilde{C}^{\alpha\beta\gamma}_{ij}}{\Lambda^2} \; q_i \bar d_j \, {\phi}_\alpha^\dagger  \phi_\beta  \phi_\gamma^{\dagger} \,,
	\end{split}
	\label{eq:HDO}
\end{align}
where the $C_{ij},\, \tilde{C}_{ij}$ tensors respect Higgs-flavor and CP symmetries. These operators generate corrections to $x$ and $\tilde{x}$ Yukawa interactions at one-loop by 
connecting two scalar legs, as shown in Fig.~\ref{fig:dim6}. For the diagram on the left, the loop integral is quadratically divergent, corresponding to a threshold correction at scale $\Lambda$. However, this contribution preserves the Higgs-flavor symmetry, and only corrects the size of non-zero entries of $x,\, \tilde{x}$, which we take to be free parameters. Introducing the flavor and CP violating scalar mass terms $\mu^2_{\beta\gamma}$, represented by a cross in the diagram on the right in Fig.~\ref{fig:dim6}, leads to complex contributions to elements of $x$ and $\tilde{x}$ (including those that vanish at tree-level), and hence can give a contribution to $\bar\theta$. With the $\mu^2_{\beta \gamma}$ insertion, the loop integral is logarithmically divergent, so that these flavor-violating contributions are generated by RG scaling from the scale $\Lambda$ to $M$ and are proportional to $\ln (\Lambda/M$)
\begin{align}
	\begin{split}
		\delta x^\alpha_{ij} &= \frac{1}{8 \pi^2 \Lambda^2} \ln \bigg(\frac{\Lambda}{M}\bigg) \sum_{\beta \gamma} C^{\alpha\beta\gamma}_{ij} \; \mu^2_{\beta\gamma}\,,\quad
		\delta \tilde{x}^\alpha_{ij} = \frac{1}{8 \pi^2 \Lambda^2} \ln \bigg(\frac{\Lambda}{M}\bigg) \sum_{\beta \gamma} \tilde{C}^{\alpha\beta\gamma}_{ij} \; \mu^2_{\beta\gamma}.
	\end{split}
\end{align}
We ignore the threshold corrections at scale $M$, which have constant terms and terms proportional to logs of heavy Higgs mass ratios.
\begin{figure}[H]
	\centering
	\begin{tikzpicture}
		\begin{feynman}
			\vertex (L) at (-2,0);
			\vertex (R) at (+2,0);
			\vertex (M) at (0,0);
			\vertex (lab) at (0,2.5) {$\phi_\beta$};
			\vertex (B) at (0,-1.5) {$\phi_\alpha$};
			
			\diagram*{
				(M) -- [anti fermion, edge label'={$\bar d_j$ or $\bar u_j$}] (R),
				(L) -- [fermion, edge label'={$q_i$}] (M),
				(M) -- [scalar] (B),
			};
			\draw[charged scalar] (M) arc [start angle = 270, end angle = -90, radius = 1];
		\end{feynman}
	\end{tikzpicture}
	\qquad
	\begin{tikzpicture}
		\begin{feynman}
			\vertex (L) at (-2,0);
			\vertex (R) at (+2,0);
			\vertex (M) at (0,0);
			\vertex (T) at (0,2);
			\vertex at (0,2.5) {$\mu^2_{\gamma\beta}$};
			\vertex (B) at (0,-1.5) {$\phi_\alpha$};
			
			\diagram*{
				(M) -- [anti fermion, edge label'={$\bar{d}_j$ or $\bar u_j$}] (R),
				(L) -- [fermion, edge label' = {$q_i$}] (M),
				(M) -- [scalar] (B),
				(M) -- [charged scalar, half left, looseness=1.7, insertion = 0.99, edge label={$\phi_\beta$}] (T),
				(T) -- [charged scalar, half left,looseness=1.7,edge label = {$\phi_\gamma$}] (M),
			};
		\end{feynman}
	\end{tikzpicture}
	\caption{One-loop radiative corrections to the $x_{ij}^\alpha,\tilde{x}_{ij}^\alpha$ matrices from dimension six operators and soft scalar masses in the interaction basis.}
	\label{fig:dim6}
\end{figure}

The resulting radiative contribution to $\bar{\theta}$ is obtained by inserting these corrections into $y^u$ and $y^d$. A precise numerical evaluation is not possible, because  $C^{\alpha\beta\gamma}_{ij}$ and $\tilde{C}^{\alpha\beta\gamma}_{ij}$ are unknown.  However, using approximate $\rm U(1)^9$ flavor symmetries the non-zero entries are estimated to be $C_{ij} \sim \epsilon_{q_i} \epsilon_{u_j}, \, \tilde{C}_{ij} \sim \epsilon_{q_i} \epsilon_{d_j} $. Therefore, the magnitudes of all contributions to the determinant of $y^u$ and $y^d$, whether at tree or loop level, are proportional to $\epsilon_{q_1} \epsilon_{q_2} \epsilon_{q_3} \epsilon_{u_1} \epsilon_{u_2} \epsilon_{u_3}$  and $\epsilon_{q_1} \epsilon_{q_2} \epsilon_{q_3} \epsilon_{d_1} \epsilon_{d_2} \epsilon_{d_3}$, respectively, which drops out of the result for $\bar{\theta}$, giving the rough estimate 
\begin{align}
	\delta \bar{\theta}_{\Lambda} \simeq  \frac{1}{8 \pi^2} \frac{M^2}{\Lambda^2} \ln \bigg(\frac{\Lambda}{M}\bigg) \sum_{\beta\gamma}\; \frac{\mu^2_{\gamma\beta}}{M^2} f_{\gamma\beta}(U) .
\end{align}
Here $f_{\gamma\beta}(U)$ are model-dependent cubic polynomials of matrix elements of $U$. The off-diagonal matrix elements of $U$ are small if the scale of the soft parameters $\mu^2_{\alpha \beta}$
are less than the mass scale $M$ of the heavy Higgs. While the suppression from this is limited, since realistic quark masses and mixings must result from perturbative values of the $x$ and $\tilde{x}$ matrices, the general expectation is that $\Lambda$ need not be many orders of magnitude above $M$. We conclude by discussing higher-dimensional operators in the two models presented in the main letter.

\subsection{N=2}

For the model with two Higgs doublets, the contributions to $\bar{\theta}$  at $\mathcal{O}\big(\frac{1}{\Lambda^2}\big)$, from one loop diagrams involving dimension 6 operators, read
\begin{align}
	\begin{split}
		\delta\bar{\theta}_{\Lambda} \simeq \bigg[& -2\, \tilde{C}^{212}_{31}\, \sin\xi\, +3408\,\tilde{C}^{212}_{32}\, \sin^2\xi\,\tilde{x}_{21} + 10\, \tilde{C}_{31}^{212}\cos\xi \sin\xi\, x_{32} \\
		&-2\times 10^4\, \tilde{C}_{32}^{212}\cos\xi \, \sin^2\xi\, \tilde{x}_{21} x_{32} -(7\times 10^3\, C_{12}^{212} - 10^3\, C_{13}^{212})\sin^2\xi\, x_{31}\\
		& + (5\times 10^7 C_{12}^{212} - 8\times 10^6 C_{13}^{212})\sin^3\xi\, \tilde{x}_{21}\, x_{31}\bigg]\frac{\mu^2}{\Lambda^2}\log\bigg( \frac{\Lambda^2}{M^2}\bigg)\sin3\beta\,.
	\end{split}
\end{align}
Using a $\rm U(1)^9$ flavor symmetry, we find $\sin\xi\sim 0.4$ and $M/\Lambda \lesssim 8\times 10^{-5}$ to not spoil the solution to the strong CP problem.

\subsection{N=3}

For the model with 3 Higgs doublets, the contributions to $\bar{\theta}$  at $\mathcal{O}\big(\frac{1}{\Lambda^2}\big)$, from one loop diagrams involving dimension 6 operators, read
\begin{align}
	\begin{split}
		\delta\bar{\theta}_{\Lambda} \simeq&\; \frac{M^2}{\Lambda^2}\log\bigg(\frac{\Lambda^2}{M^2}\bigg)\bigg[ \big(4\times 10^2\, \tilde{C}_{11}^{121}-10^3\, C_{11}^{112} \big)\epsilon_{12}\sin(\beta_{12}+2\beta_{13})\\
		&+ 4\times 10^2\, \tilde{C}_{31}^{132}\, \epsilon_{12}\,\epsilon_{13}\sin(2\beta_{12})-3\times 10^2\, C_{21}^{123}\, \epsilon_{12}^2\, \epsilon_{13}\, x_{32}\, \sin(2\beta_{12}-2\beta_{13})\\
		& + \big( 3\times 10^2\, C_{21}\, \epsilon_{12}\,\epsilon_{12}\, x_{32} - \tilde{C}_{31}^{123}\epsilon_{23}\big)\sin(\beta_{12}-\beta_{13}-\beta_{23})\bigg]\,.
	\end{split}
\end{align}
In this model we cannot use $\rm U(1)^9$ approximate flavor symmetries to estimate the sizes of non-zero entries in the $x, \tilde{x}, C, \tilde{C}$ matrices given the large size of some parameters. Instead we use our illustrative example of $\epsilon_{12} \simeq 10^{-3}, \, \epsilon_{13} \simeq 10^{-2}, \, \epsilon_{23} \simeq 10^{-5}$ and $x_{32} \simeq 10^{-2}$ which gave $\delta\bar{\theta} \simeq 10^{-10}$.  We find $\delta\bar{\theta}_{\Lambda} \lesssim 10^{-10}$ for $M/\Lambda \lesssim 10^{-3}$ when we take $C_{11}^{112}, \tilde{C}_{11}^{121} \simeq 10^{-4}$, $\tilde{C}_{31}^{132} \simeq 10^{-2}$ and other $C, \tilde{C}$ elements of order unity.

\section{Spontaneous CP breaking}

In most of this paper we treat $\mu^2_{\alpha \beta}$ as soft symmetry-breaking spurions for $G_{\rm HF}$ and $CP$. However, they may arise from spontaneous $G_{\rm HF}$ and $CP$ symmetry breaking via a set of SM singlet vacuum expectation values (vevs), {$\xi$}. The dimension of the operator linking the Higgs bilinear $\phi^\dagger_\alpha \phi_\beta$ to the $\xi$ fields depends on the $G_{\rm HF}$ charges of the bilinear and of $\xi$. Thus $|\mu^2_{\alpha \beta}|$, which we take less than or of order $M^2$, could be of order $M\xi, \xi^2, \xi^3/\Lambda_\phi, \xi^4/\Lambda_\phi^2, ...$, where $\Lambda_\phi$ is the UV scale at which relevant interactions appear that connect $\xi$ and $\phi_\alpha$. The various $\phi^\dagger_\alpha \phi_\beta$ operators can have different charges and the corresponding $\mu^2_{\alpha \beta}$ can differ by orders of magnitude. In the case that two such operators have the same charge, we expect the corresponding soft masses to be comparable. There could be higher dimensional interactions such as $g_1(\xi/\Lambda_q) \, q_i \bar{u}_j \phi^\alpha $ or $[g_2(\xi/\Lambda_G) - g_2^\dagger(\xi/\Lambda_G)] \, G \tilde{G}$, leading to $G_{\rm HF}$ and $CP$ breaking effects suppressed by powers of $\xi/\Lambda_{q,G}$. The polynomials $g_{1,2}$ are model-dependent, and $g_2$ is at least cubic order. We assume these can be neglected compared to breaking terms involving powers of $\xi/M$ that arise in the Higgs doublet mass matrix. In this paper, the breaking of $CP$ and $G_{\rm HF}$ is communicated to the quark sector through the Higgs Portal.
We conclude the discussion showing how the soft breaking of CP and $G_{\rm HF}$ could be obtained spontaneously in the model with 3 Higgs doubblets discussed in the main text. We introduce the scalar fields $\xi_{1,2}$ of $G_{\rm HF}$ charges $(1,2)$ that spontaneously break $G_{\rm HF}$ and CP via the vevs
\begin{align}
\begin{split}
\xi_2^* = \frac{\mu_{12}^2}{M_2} e^{i \beta_{12}} = \epsilon_{12} M_2 e^{i \beta_{12}} \simeq 10^{-3} M_2 e^{i \beta_{12}}\,,\\
\xi_1 = \frac{\mu_{13}^2}{M_3} e^{i \beta_{13}} = \epsilon_{13} M_3 e^{i \beta_{13}} \simeq 10^{-2} M_3 e^{i \beta_{13}}\,.
\end{split}
\label{eq:xiII}
\end{align}
The mass matrix for the scalars $(\phi_1, \phi_2, \phi_3)$ of $G_{\rm HF}$ charges $(0,2,-1)$ can take the form
\begin{align}
     M^2_\phi \; \simeq \; 
     \begin{pmatrix}
        M_1^2 & M_2 \xi_2^* & M_3 \xi_1 \\
        M_2 \xi_2 & M_2^2 & \xi_1 \xi_2^* \\
        M_3 \xi_1^* & \xi^*_1 \xi_2 & M_3^2 
    \end{pmatrix}
    %\hspace{0.3in}
     \; \simeq \;
    M_2^2\begin{pmatrix}
        r^2 10^{-4} &  10^{-3} e^{i \beta_{12}} & 10^{-2} r^2 e^{i \beta_{13}} \\
        10^{-3} e^{-i \beta_{12}} & 1 & 10^{-5} r e^{i (\beta_{12} + \beta_{13})}  \\
        10^{-2} r^2 e^{-i \beta_{13}} & 10^{-5} r e^{-i (\beta_{12} + \beta_{13})} & r^2
    \end{pmatrix}\,,
    \label{eq:M2phiII}
\end{align}
where $r = M_3/M_2$. Note that the 11 entry is small by the fine-tuning to obtain the weak scale much below the masses of the heavy Higgs doublets.

\renewcommand{\refname}{\hskip-1.8em References}
\bibliography{biblio}

\begin{thebibliography}{12}%
\makeatletter
\providecommand \@ifxundefined [1]{%
 \@ifx{#1\undefined}
}%
\providecommand \@ifnum [1]{%
 \ifnum #1\expandafter \@firstoftwo
 \else \expandafter \@secondoftwo
 \fi
}%
\providecommand \@ifx [1]{%
 \ifx #1\expandafter \@firstoftwo
 \else \expandafter \@secondoftwo
 \fi
}%
\providecommand \natexlab [1]{#1}%
\providecommand \enquote  [1]{``#1''}%
\providecommand \bibnamefont  [1]{#1}%
\providecommand \bibfnamefont [1]{#1}%
\providecommand \citenamefont [1]{#1}%
\providecommand \href@noop [0]{\@secondoftwo}%
\providecommand \href [0]{\begingroup \@sanitize@url \@href}%
\providecommand \@href[1]{\@@startlink{#1}\@@href}%
\providecommand \@@href[1]{\endgroup#1\@@endlink}%
\providecommand \@sanitize@url [0]{\catcode `\\12\catcode `\$12\catcode
  `\&12\catcode `\#12\catcode `\^12\catcode `\_12\catcode `\%12\relax}%
\providecommand \@@startlink[1]{}%
\providecommand \@@endlink[0]{}%
\providecommand \url  [0]{\begingroup\@sanitize@url \@url }%
\providecommand \@url [1]{\endgroup\@href {#1}{\urlprefix }}%
\providecommand \urlprefix  [0]{URL }%
\providecommand \Eprint [0]{\href }%
\providecommand \doibase [0]{https://doi.org/}%
\providecommand \selectlanguage [0]{\@gobble}%
\providecommand \bibinfo  [0]{\@secondoftwo}%
\providecommand \bibfield  [0]{\@secondoftwo}%
\providecommand \translation [1]{[#1]}%
\providecommand \BibitemOpen [0]{}%
\providecommand \bibitemStop [0]{}%
\providecommand \bibitemNoStop [0]{.\EOS\space}%
\providecommand \EOS [0]{\spacefactor3000\relax}%
\providecommand \BibitemShut  [1]{\csname bibitem#1\endcsname}%
\let\auto@bib@innerbib\@empty
%</preamble>
\bibitem [{\citenamefont {Pendlebury}\ \emph {et~al.}(2015)\citenamefont
  {Pendlebury} \emph {et~al.}}]{Pendlebury:2015lrz}%
  \BibitemOpen
  \bibfield  {author} {\bibinfo {author} {\bibfnamefont {J.~M.}\ \bibnamefont
  {Pendlebury}} \emph {et~al.},\ }\bibfield  {title} {\bibinfo {title}
  {{Revised experimental upper limit on the electric dipole moment of the
  neutron}},\ }\href {https://doi.org/10.1103/PhysRevD.92.092003} {\bibfield
  {journal} {\bibinfo  {journal} {Phys. Rev. D}\ }\textbf {\bibinfo {volume}
  {92}},\ \bibinfo {pages} {092003} (\bibinfo {year} {2015})},\ \Eprint
  {https://arxiv.org/abs/1509.04411} {arXiv:1509.04411 [hep-ex]} \BibitemShut
  {NoStop}%
\bibitem [{\citenamefont {Nelson}(1984)}]{Nelson:1983zb}%
  \BibitemOpen
  \bibfield  {author} {\bibinfo {author} {\bibfnamefont {A.~E.}\ \bibnamefont
  {Nelson}},\ }\bibfield  {title} {\bibinfo {title} {{Naturally Weak CP
  Violation}},\ }\href {https://doi.org/10.1016/0370-2693(84)92025-2}
  {\bibfield  {journal} {\bibinfo  {journal} {Phys. Lett. B}\ }\textbf
  {\bibinfo {volume} {136}},\ \bibinfo {pages} {387} (\bibinfo {year}
  {1984})}\BibitemShut {NoStop}%
\bibitem [{\citenamefont {Barr}(1984)}]{Barr:1984qx}%
  \BibitemOpen
  \bibfield  {author} {\bibinfo {author} {\bibfnamefont {S.~M.}\ \bibnamefont
  {Barr}},\ }\bibfield  {title} {\bibinfo {title} {{Solving the Strong CP
  Problem Without the Peccei-Quinn Symmetry}},\ }\href
  {https://doi.org/10.1103/PhysRevLett.53.329} {\bibfield  {journal} {\bibinfo
  {journal} {Phys. Rev. Lett.}\ }\textbf {\bibinfo {volume} {53}},\ \bibinfo
  {pages} {329} (\bibinfo {year} {1984})}\BibitemShut {NoStop}%
\bibitem [{\citenamefont {Hiller}\ and\ \citenamefont
  {Schmaltz}(2002)}]{Hiller:2002um}%
  \BibitemOpen
  \bibfield  {author} {\bibinfo {author} {\bibfnamefont {G.}~\bibnamefont
  {Hiller}}\ and\ \bibinfo {author} {\bibfnamefont {M.}~\bibnamefont
  {Schmaltz}},\ }\bibfield  {title} {\bibinfo {title} {{Strong Weak CP
  Hierarchy from Nonrenormalization Theorems}},\ }\href
  {https://doi.org/10.1103/PhysRevD.65.096009} {\bibfield  {journal} {\bibinfo
  {journal} {Phys. Rev. D}\ }\textbf {\bibinfo {volume} {65}},\ \bibinfo
  {pages} {096009} (\bibinfo {year} {2002})},\ \Eprint
  {https://arxiv.org/abs/hep-ph/0201251} {arXiv:hep-ph/0201251} \BibitemShut
  {NoStop}%
\bibitem [{\citenamefont {Feruglio}\ \emph {et~al.}(2023)\citenamefont
  {Feruglio}, \citenamefont {Strumia},\ and\ \citenamefont
  {Titov}}]{Feruglio:2023uof}%
  \BibitemOpen
  \bibfield  {author} {\bibinfo {author} {\bibfnamefont {F.}~\bibnamefont
  {Feruglio}}, \bibinfo {author} {\bibfnamefont {A.}~\bibnamefont {Strumia}},\
  and\ \bibinfo {author} {\bibfnamefont {A.}~\bibnamefont {Titov}},\ }\bibfield
   {title} {\bibinfo {title} {{Modular invariance and the QCD angle}},\ }\href
  {https://doi.org/10.1007/JHEP07(2023)027} {\bibfield  {journal} {\bibinfo
  {journal} {JHEP}\ }\textbf {\bibinfo {volume} {07}},\ \bibinfo {pages}
  {027}},\ \Eprint {https://arxiv.org/abs/2305.08908} {arXiv:2305.08908
  [hep-ph]} \BibitemShut {NoStop}%
\bibitem [{\citenamefont {Feruglio}\ \emph {et~al.}(2024)\citenamefont
  {Feruglio}, \citenamefont {Parriciatu}, \citenamefont {Strumia},\ and\
  \citenamefont {Titov}}]{Feruglio:2024ytl}%
  \BibitemOpen
  \bibfield  {author} {\bibinfo {author} {\bibfnamefont {F.}~\bibnamefont
  {Feruglio}}, \bibinfo {author} {\bibfnamefont {M.}~\bibnamefont
  {Parriciatu}}, \bibinfo {author} {\bibfnamefont {A.}~\bibnamefont
  {Strumia}},\ and\ \bibinfo {author} {\bibfnamefont {A.}~\bibnamefont
  {Titov}},\ }\bibfield  {title} {\bibinfo {title} {{Solving the strong CP
  problem without axions}},\ }\href@noop {} {\  (\bibinfo {year} {2024})},\
  \Eprint {https://arxiv.org/abs/2406.01689} {arXiv:2406.01689 [hep-ph]}
  \BibitemShut {NoStop}%
\bibitem [{\citenamefont {Penedo}\ and\ \citenamefont
  {Petcov}(2024)}]{Penedo:2024gtb}%
  \BibitemOpen
  \bibfield  {author} {\bibinfo {author} {\bibfnamefont {J.~T.}\ \bibnamefont
  {Penedo}}\ and\ \bibinfo {author} {\bibfnamefont {S.~T.}\ \bibnamefont
  {Petcov}},\ }\bibfield  {title} {\bibinfo {title} {{Finite modular symmetries
  and the strong CP problem}},\ }\href@noop {} {\  (\bibinfo {year} {2024})},\
  \Eprint {https://arxiv.org/abs/2404.08032} {arXiv:2404.08032 [hep-ph]}
  \BibitemShut {NoStop}%
\bibitem [{\citenamefont {Georgi}(1978)}]{Georgi:1978xz}%
  \BibitemOpen
  \bibfield  {author} {\bibinfo {author} {\bibfnamefont {H.}~\bibnamefont
  {Georgi}},\ }\bibfield  {title} {\bibinfo {title} {{A Model of Soft CP
  Violation}},\ }\href@noop {} {\bibfield  {journal} {\bibinfo  {journal}
  {Hadronic J.}\ }\textbf {\bibinfo {volume} {1}},\ \bibinfo {pages} {155}
  (\bibinfo {year} {1978})}\BibitemShut {NoStop}%
\bibitem [{\citenamefont {Mohapatra}\ and\ \citenamefont
  {Wyler}(1980)}]{Mohapatra:1979kg}%
  \BibitemOpen
  \bibfield  {author} {\bibinfo {author} {\bibfnamefont {R.~N.}\ \bibnamefont
  {Mohapatra}}\ and\ \bibinfo {author} {\bibfnamefont {D.}~\bibnamefont
  {Wyler}},\ }\bibfield  {title} {\bibinfo {title} {{A Solution to the Strong
  {CP} Problem in the SU(5) Model}},\ }\href
  {https://doi.org/10.1016/0370-2693(80)90005-2} {\bibfield  {journal}
  {\bibinfo  {journal} {Phys. Lett. B}\ }\textbf {\bibinfo {volume} {89}},\
  \bibinfo {pages} {181} (\bibinfo {year} {1980})}\BibitemShut {NoStop}%
\bibitem [{\citenamefont {Glashow}(2001)}]{Glashow:2001yz}%
  \BibitemOpen
  \bibfield  {author} {\bibinfo {author} {\bibfnamefont {S.~L.}\ \bibnamefont
  {Glashow}},\ }\bibfield  {title} {\bibinfo {title} {{A Simple solution to the
  strong CP problem}},\ }\href@noop {} {\  (\bibinfo {year} {2001})},\ \Eprint
  {https://arxiv.org/abs/hep-ph/0110178} {arXiv:hep-ph/0110178} \BibitemShut
  {NoStop}%
\bibitem [{\citenamefont {Ellis}\ and\ \citenamefont
  {Gaillard}(1979)}]{Ellis:1978hq}%
  \BibitemOpen
  \bibfield  {author} {\bibinfo {author} {\bibfnamefont {J.~R.}\ \bibnamefont
  {Ellis}}\ and\ \bibinfo {author} {\bibfnamefont {M.~K.}\ \bibnamefont
  {Gaillard}},\ }\bibfield  {title} {\bibinfo {title} {{Strong and Weak CP
  Violation}},\ }\href {https://doi.org/10.1016/0550-3213(79)90297-9}
  {\bibfield  {journal} {\bibinfo  {journal} {Nucl. Phys. B}\ }\textbf
  {\bibinfo {volume} {150}},\ \bibinfo {pages} {141} (\bibinfo {year}
  {1979})}\BibitemShut {NoStop}%
\bibitem [{\citenamefont {Ferreira}\ and\ \citenamefont
  {Lavoura}(2019)}]{Ferreira:2019aps}%
  \BibitemOpen
  \bibfield  {author} {\bibinfo {author} {\bibfnamefont {P.~M.}\ \bibnamefont
  {Ferreira}}\ and\ \bibinfo {author} {\bibfnamefont {L.}~\bibnamefont
  {Lavoura}},\ }\bibfield  {title} {\bibinfo {title} {{No strong $CP$ violation
  up to the one-loop level in a two-Higgs-doublet model}},\ }\href
  {https://doi.org/10.1140/epjc/s10052-019-7053-4} {\bibfield  {journal}
  {\bibinfo  {journal} {Eur. Phys. J. C}\ }\textbf {\bibinfo {volume} {79}},\
  \bibinfo {pages} {552} (\bibinfo {year} {2019})},\ \Eprint
  {https://arxiv.org/abs/1904.08438} {arXiv:1904.08438 [hep-ph]} \BibitemShut
  {NoStop}%
\end{thebibliography}%

\end{document}